\documentclass[pre,aps,preprint,nofootinbib,longbibliography]{revtex4-1}
\usepackage{amsmath}
\usepackage{amssymb}
\usepackage{graphicx}
\usepackage{hyperref}
\usepackage{array}
\usepackage{physics}
\usepackage{amssymb}
\begin{document}

\date{\today }
\title{Investigation of on-axis and off-axis levitation by a rotating permanent magnet}
\author{Hugo Schreckenberg}
\email{hugoschreckenberg@gmail.com}
\author{Zayneb El Omari El Alaoui}
\affiliation{Ecole polytechnique, Institut Polytechnique de Paris, Palaiseau, France}
\author{Guilhem Gallot}
\email{Guilhem.Gallot@polytechnique.edu}
\affiliation{LOB, CNRS, INSERM, Ecole polytechnique, Institut Polytechnique de Paris, Palaiseau, France}

\begin{abstract}
A slightly tilted permanent magnet rotating at high speed can induce a magnetic field capable of trapping another permanent magnet in a gravity independent levitated bound state, bypassing Earnshaw's theorem. During levitation, the floater magnet is locked in a conical orbit at the same frequency as the rotor. This rotation allows the sides of the same polarity of each magnet to face each other, which is responsible for the dynamic equilibrium of the floater magnet. Here, we theoretically explain the motion of the floater in-axis and off-axis and highlight levitation stability conditions and their dependence on the size of the floater and the speed of the rotor. We also experimentally studied the levitation conditions with respect to the rotational speed of the rotor for various floater's sizes and shapes. We observed and analyzed the lower and upper limits of levitation. Finally, we explained the off-axis motion of the center of mass of the floater from its equilibrium position by an extension of the dipole moment model. 
\end{abstract}
\maketitle

\section{Introduction}

Magnetic levitation could turn our science-fiction dreams into reality. Numerous ways of bypassing Earnshaw's electrostatic theorem \cite{griffiths2023introduction}, which prohibits levitation with magnets in many common situations, have been discovered. Diamagnetic materials can levitate above permanent magnets \cite{simon2000diamagnetic}. At low temperatures, levitation is possible with superconductors thanks to the Meissner effect \cite{kim2019levitation}. Spin-stabilized magnetic levitation circumvents Earnshaw's theorem by using the rapid rotation of a magnet \cite{ michaelis2020variety, simon1997spin}. The main applications of magnetic levitation are magnetic levitation trains \cite{lee2006review}, magnetic bearings \cite{schweitzer2009magnetic}, flywheels \cite{supreeth2022review}, microbotics \cite{shameli2008design}, contactless motor or inertial sensing \cite{wangDynamicsFerromagneticParticle2019,timberlakeAccelerationSensingMagnetically2019}.

A new type of levitation using rotating permanent magnets has recently been discovered by Ucar \cite{ucar2021polarity}. A first permanent magnet, called the rotor, rotates at high speed with its magnetic moment almost perpendicular to the axis of rotation. A second magnet, the floater, is placed on the axis of rotation, above or below the rotor. The floater follows a conical trajectory at the same frequency as the rotor and enters into a bound state with the rotor.  In his pioneering paper \cite{ucar2021polarity}, Ucar presents the floating motion in different configurations. Additional studies on the angles of rotation were conducted by Le Lay \textit{et al} \cite{le2024magnetic} and Hermansen \textit{et al} \cite{hermansen2023magnetic, hermansenMagneticLevitationLow2025}. The latter also improved the understanding of the influence of the size and magnetization of the floater and of the time stability of the bound state in different configurations. They also addressed the issue of the out-of-equilibrium motion of the floater, showing multiple modes without providing a precise physical explanation. In this paper, we provide a thorough theory of the rotor/floater bound system and extend it to the case where the floater is laterally out of equilibrium. We theoretically refine the equilibrium conditions of the floater levitation, which compare well with the experimental data.
This article is structured as follows: Part II presents the theoretical framework, and Part III presents the experimental results. More precisely, Section II.A defines the problem and the general equations for the levitation. Section II.B examines the vertical equilibrium and Section II.C the lateral equilibrium when the floater remains on the axis of rotation of the rotor. Section II.D studies the center of mass out of equilibrium, finally, Part III shows and compares the experimental results with the theory.

\section{Theory}
\subsection{General considerations}

\begin{figure}[tb]
    \centering
   \includegraphics[scale= 1]{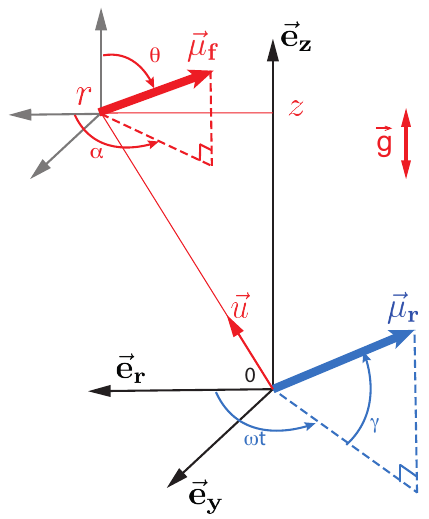} 
    \caption{Three-dimensional description in the $(\vb{e_r},\vb{e_y},\vb{e_z})$ basis of the positions of the rotor (magnetic moment vector in blue) and of the floater in levitation (magnetic moment vector in red).} 
    \label{euler}
\end{figure}

The axis of rotation of the rotor is denoted $\vb{e_z}$ (see Fig.~\ref{euler}). The unit vector connecting the centers of gravity of the two magnets is given by $\vb{u}$. The unit vector perpendicular to $\vb{e_z}$ in the plane $(\vb{e_z},\vb{u})$ is denoted $\vb{e_r}$, and the unit vector $\vb{e_y}$ is such that $(\vb{e_r},\vb{e_y},\vb{e_z})$ is a direct orthonormal basis of the space. The origin 0 is the center of the rotor magnet. In this system of coordinates, $z$ is the height of the floater and $r$ its deviation from the rotational axis, so $\vb{u} = (r,0,z)/\sqrt{z^2+r^2}$. The following study is made in this frame of reference. The rotor and floater magnets are considered to be magnetic dipoles, respectively $\boldsymbol{\mu_r}$ and $\boldsymbol{\mu_f}$. An important parameter for levitation is the angle $\gamma$ between the moment of the rotor $\boldsymbol{\mu_r}$ and the plane $(\vb{e_r},\vb{e_y})$.

Since the floater is treated as a dipole, its rotation around its magnetic moment is not considered in the present model. Therefore, for a fixed center of mass, the orientation of the magnetic moment of the floater is determined by its polar angle $\theta$ with respect to axis $\vb{e_z}$, and its azimuthal angle $\alpha$ with $\vb{e_r}$ in spherical coordinates. Thus, the rotor and floater magnetic moments can be expressed as follows
\begin{equation}\label{mR}
    \boldsymbol{\mu_r} = \mu_r \begin{pmatrix}
        \cos(\omega t)\cos\gamma \\ 
        \sin(\omega t)\cos\gamma \\ 
        \sin\gamma 
    \end{pmatrix}
\qquad
 \boldsymbol{\mu_f} = \mu_f \begin{pmatrix}
        \cos\alpha \sin\theta \\ 
        \sin\alpha \sin\theta \\ 
        \cos\theta
    \end{pmatrix}
\end{equation}
where $\omega$ is the imposed angular frequency of the rotor. Using the following basic magnetostatic equations for the magnetic field $\vb{B_d}$ generated by the rotor and for the potential magnetic energy $E_m$ of the floater \cite{seleznyova2016modelling, greene1971force}
\begin{equation}
   \vb{B_d} = \frac{\mu_0}{4\pi}\,\frac{3(\boldsymbol{\mu_r}\cdot \vb{u})\vb{u}- \boldsymbol{\mu_r}}{(z^2+r^2)^{3/2}},
\qquad
     E_m = - \boldsymbol{\mu_f} \cdot \vb{B_d},
\end{equation}
the total potential energy of the floater is derived as the sum of the magnetic $E_m$ and gravity $E_g$ terms
\begin{align}
E_p &= E_m + E_g \nonumber\\
& =  -\frac{\mu_0 \mu_r \mu_f}{4\pi(r^2+z^2)^{\frac{5}{2}}} \Big\{  z^2 \big[ 2\cos\theta \sin\gamma - \sin\theta \cos\gamma \cos(\omega t - \alpha ) \big] \nonumber \\
& \quad + 3rz \big[ \sin \gamma \sin\theta \cos\alpha + \cos\theta\cos\gamma \cos(\omega t )\big]  \nonumber \\
& \quad +  r^2 \big[  \cos\gamma \sin\theta \; [2 \cos\alpha\cos(\omega t) - \sin\alpha \sin(\omega t)] - \sin\gamma \cos\theta \big] \Big\} \pm mgz. \label{genformula1}
\end{align}

Here, the floater is considered to be affected only by the external magnetic and gravitational fields. The notation $\pm$ in Eq.~\eqref{genformula1} includes both the situation when the floater is above the rotor (+) and when the floater is below the rotor (-). The difference between the two positions is that gravity pulls the floater towards the rotor when it is above, and pushes it away when it is below. In the following, both notations $\pm$ and $\mp$ will be used. The sign at the top corresponds to the situation when the floater is above the rotor, the sign at the bottom corresponds to the situation when the floater is below the rotor. Equation \ref{genformula1} can be generalized to the case where the axis of rotation is tilted by an angle $\beta$ by replacing $\pm mgz$ with $mg[z \cos\beta + r\sin\beta]$.
A careful examination of the motion of the magnet in slow motion \cite{slowmo_ucar, slowmo_hugo} reveals a remarkable observation: the magnetic moment of the floating magnet moves at the same frequency as the rotor whereas the center of mass displacement is negligible over a rotation. Thus, considering the latter fixed over a rotation, 
the orientation of the floater follows the laws of a spherical pendulum described in \cite{richter1996spherical} by
\begin{equation}
\mathcal{L} = I\left[ \dot \theta ^2 + \dot \alpha^2 \sin^2\theta\right] - E_p,
\end{equation}
\begin{equation}
\frac{d}{dt}\frac{\partial}{\partial \dot \theta }\mathcal{L} - \frac{\partial}{\partial \theta } \mathcal{L} = 0, \label{lagphi}
\end{equation}
\begin{equation}
\frac{d}{dt}\frac{\partial}{\partial \dot \alpha }\mathcal{L} - \frac{\partial}{\partial \alpha } \mathcal{L} = 0, \label{lagalpha}
\end{equation}
where $\mathcal{L}$ is the Lagrangian and $I$ is the moment of inertia of the floater taken at its center, along the direction of its magnetic moment. The floater is supposed magnetically homogeneous; therefore its center of mass is at the same position as that of the magnetic dipole. Thus, the torque of gravity is not taken into account. These equations give the coupled differential equations 
\begin{align}
2 I\frac{d^2 \theta}{dt^2} &= 2I \left( \frac{d\alpha}{dt} \right)^2 \sin\theta \cos\theta + \frac{\mu_0 \mu_r \mu_f}{4 \pi (z^2 + r^2)^{5/2}} \Big\{ -z^2[2\sin\theta \sin\gamma + \cos\theta\cos\gamma\cos(\omega t - \alpha)] \nonumber \\ 
& \quad + 3rz[\cos\theta \sin\gamma \cos\alpha - \sin\theta\cos\gamma\cos(\omega t )] \nonumber \\
& \quad + r^2\big[\sin\gamma \sin\theta + \cos\gamma \cos\theta [2\cos\alpha\cos(\omega t) - \sin\alpha\sin(\omega t)] \big]\Big\} \label{lagphi2}
\end{align}
\begin{align}
2I \frac{d}{dt}\left[ \frac{d\alpha}{dt} \sin^2\theta \right]  & = \frac{\mu_0 \mu_r \mu_f}{4 \pi (z^2 + r^2)^{5/2}} \Big\{ - z^2 \sin\theta \cos\gamma \sin(\omega t - \alpha) - 3rz \sin\gamma\sin\theta\sin\alpha \nonumber \\
&  \quad -r^2 \cos\gamma \sin\theta \big[ 2\sin\alpha\cos(\omega t) + \cos\alpha\sin(\omega t)\big] \Big\}.
\label{lagphi3}
\end{align}

\subsection{Conical motion and stable equilibrium along the axis of rotation} \label{conical}

Placed along the axis of rotation, the motion of the magnetic moment of the floater can be accurately described as a conical trajectory of angle $\theta$ constant around the $z$-axis. This configuration corresponds to the case where $\alpha = \omega t$, which allows poles of the same polarity to face each other. This configuration cannot happen in a static case and is allowed here thanks to the dynamics of the system. 


For $r = 0$, a solution for Eqs.~\eqref{lagphi} and \eqref{lagalpha} is found for $\theta$ constant and $ \alpha = \omega t $. Equation~\eqref{lagphi2} then determines the value of $\theta$ by
\begin{equation}\label{phi}
    \frac{2\sin\gamma}{\cos\theta} + \frac{\cos\gamma}{\sin\theta} = \frac{8\pi I\omega^2 z^3}{\mu_0 \mu_r \mu_f}.
\end{equation} 
It follows that in the limit of small angles $\gamma \ll 1$,  $\sin \theta$ increases with the inverse of $\omega^2$. 

For Eq. \eqref{phi} to have a solution for $\theta$ , it requires the right-hand side of Eq. \eqref{phi} to exceed the minimum value of the left-hand side. We can then analytically deduce that the angular velocity must be greater than a lower limit
$\omega_0$, which is found to be (details are provided in Appendix A)
\begin{equation}
    \omega_0^2  = \frac{\mu_0 \mu_r \mu_f}{8\pi I z^3 } \cos \gamma \Big[ 1 + (2\tan \gamma)^{2/3} \Big]^{3/2} .
    \label{omega0}
\end{equation}
In the small angle limit for $\gamma$, this lower limit $\omega_0 $ corresponds to $\theta \simeq 90^{\circ}$. Experimentally, this extreme value of $\theta$ is never observed, and $\theta$ is rarely above $20^{\circ}$. Therefore, $\omega_0$ is an underestimate of the minimal value of levitation. Replacing $\alpha$ in Eq.~\eqref{genformula1} for $r = 0$, we get a new expression for $E_p$.
\begin{align}
E_p(r = 0, z) & =  -\frac{\mu_0 \mu_r \mu_f}{4\pi z^3} \left(2 \cos \theta \sin\gamma - \cos\gamma \sin\theta \right)  \pm mgz.\label{genformula2}
\end{align}
The energy is no longer time-dependent. The vertical force applied to the floater $ F_z $ is obtained using $F_z = - \frac{\partial}{\partial_z} E_p$,
\begin{align} 
  F_z(r=0,z)   &= \mp mg + \frac{ 3 \mu_0 \mu_r \mu_f}{4\pi z^4}  \left( \cos\gamma \sin\theta - 2 \cos\theta \sin\gamma \right) \nonumber \\
  & - \frac{  \mu_0 \mu_r \mu_f}{4\pi z^3}  \left( \cos\gamma \cos\theta + 2 \sin\gamma \sin\theta \right) \frac{\partial}{\partial z} \theta.
 \label{horizontaleq}
\end{align}
Using Eq.~\eqref{phi}, $ F_z(r=0,z) $ writes
\begin{align} 
  F_z  (r=0,z)   &= \mp mg + \frac{ 3 \mu_0 \mu_r \mu_f}{4\pi z^4} \left( \cos\gamma \sin\theta - 2 \cos\theta\sin\gamma \right) \nonumber \\
  & \quad + \frac{ 6 I \omega^2}{z}  \frac{\left(1 + 2 \tan\gamma\; \tan\theta\right)\sin^2\theta}{1 - \tan\gamma\; \tan^3\theta}. \label{Fz_exact}
\end{align}

Experimentally, neither $\gamma$ nor $\theta$ exceed $45^{\circ}$. Thus, the third term of Eq.~\eqref{Fz_exact} is always positive. Consequently, when the floater is below the rotor, the first and last terms are positive. Therefore, the second term must be negative for equilibrium to be achieved. This leads to the necessary condition that
\begin{equation}
    \tan \theta < 2 \tan \gamma. \label{condition1}
\end{equation} 
For this condition to be possible, $\gamma > 0$ is mandatory. This angle is therefore essential for dynamic levitation phenomena. This condition induces a new lower bound $\omega_1$ for the angular velocity. For values of $\theta \ll 1$, Eq.~\eqref{phi} simplifies as
\begin{equation}
\tan\theta \approx \sin\theta \approx  \frac{\mu_0 \mu_r \mu_f \cos\gamma}{8\pi I \omega^2 z^3 - 2 \mu_0 \mu_r \mu_f \sin\gamma} 
\label{phi_1_order}
\end{equation}
and one obtains using Eq.~\eqref{condition1} the stability condition for
\begin{equation} 
    \omega \geq  \omega_1, 
    \qquad 
    \omega_1 = \sqrt{\frac{\mu_0 \mu_r \mu_f }{4 \pi I z^3 } \left( \frac{\cos^2\gamma}{4\sin\gamma} + \sin\gamma \right) }. \label{omega1}
\end{equation} 

This new lower bound is significantly larger than the previous one ( $\omega_{1} > \omega_0$ ) especially since $\gamma \ll 1$. Ucar, in his paper \cite{ucar2021polarity} Section 5.1.1 equation (49), gives an empirical relation for the stability of the conical motion equivalent to this one.

The effect of the weight can help or hinder the levitation. When the floater is below the rotor, the greater the gravitational force, the greater $2 \tan \gamma - \tan\theta$ and thus the greater $\omega $ has to be. The minimum angular velocity required for levitation will be greater when the floater is below the rotor. 

At high speed, the condition $\theta \ll 1$ leads to the first-order approximation for the vertical force along the axis of rotation
\begin{equation} \label{smallangle}
 F_z (r=0,z) \approx \mp mg - \frac{ 3 \mu_0 \mu_r \mu_f}{2\pi z^4}\sin\gamma + \frac{ 3\left( \mu_0 \mu_r \mu_f \cos\gamma \right)^2 }{ 16 \pi^2 I \omega^2 z^7}.  
\end{equation}

This equation emphasizes again the importance of the angle $\gamma$. A nonzero value of $\gamma$ creates an attractive force proportional to $z^{-4} $, which opposes the repulsive force proportional to $z^{-7} $, so that the floater enters a bounded state and a stable equilibrium. For $z$ small, the repulsive force is stronger, so when the floater is above the rotor, gravity helps the attractive force to maintain a bound state. This shows that levitation should be more stable in this configuration. On the other hand, when the floater is below the rotor, there is not always an equilibrium. In this case, we can prove that there exists a new lower limit $\omega_2$, under which there is no equilibrium. This new limit is only valid when the floater is below the rotor and contrary to $\omega_0$ and $\omega_1$ it depends on gravity. 
Eq.~\eqref{smallangle} is a polynomial of degree 7 in $1/z$ so there is no analytical formula for $\omega_2$. However, it is also a polynomial equation of degree 2 in $\mu_0 \mu_r \mu_f$. Thus, if $F_z (r=0,z)$ admits a zero for $z = z_{eq}$ then the discriminant of the latter polynomial should be positive: 

\begin{equation}
\label{Discriminant}
\Delta = \left(\frac{ 3 }{2\pi z_{eq}^4} \sin\gamma \right)^2 - \frac{ 3 \cos^2\gamma }{ 4 \pi^2 I \omega^2 z_{eq}^7} mg \geq 0 , 
\end{equation}
which is equivalent to
\begin{equation}
\label{omega2prime}
\omega \geq \frac{1}{\tan\gamma} \sqrt{\frac{mgz}{3I}} = \omega_2'.
\end{equation}
We can notice that we have $\omega_2 \geq \omega_2'$. 

Eq.~\eqref{smallangle} also shows that the equilibrium height decreases with the speed of the rotor. For $z$ small enough, one could neglect the weight in front of the magnetic force. This approximation also gives the following relationship for the equilibrium distance $z_{eq}$,
\begin{equation}
z_{eq}^3 = \frac{\mu_0 \mu_r \mu_f \cos^2\gamma}{8\pi I \omega^2 \sin \gamma}. \label{zeq}
\end{equation}

This result explains the experimental results from \cite{le2024magnetic} III.B, Fig. 8, which shows a better fit of $z \propto \omega^{-2/3} $ when the rotational speed is greater. Under the same assumptions, re-injecting this equation in Eq. \eqref{phi} gives
\begin{equation}
\theta \approx \frac{\cos \gamma \sin\gamma}{\cos^2\gamma - 2 \sin^2\gamma}.
\end{equation}
Thus, at equilibrium, $\theta $ would be independent on $\omega$. We recall that to have this expression $\theta \ll 1 $ has been assumed, thus the expression above can also be simplified to $ \theta \approx \gamma $ which concurs with theoretical and experimental results in \cite{le2024magnetic}. We can thus deduce an explicit formula for $\theta$ from Eq. \eqref{phi}
\begin{equation}
\theta = \frac{\mu_0 \mu_r \mu_r }{8 \pi I \omega^2 z^3},
\end{equation}
which simplifies Eq.\eqref{Fz_exact} to
\begin{equation}
F_z(r=0, z ) = \mp mg + \frac{3\mu_0 \mu_r \mu_r }{2\pi z^4}(\theta - \gamma).
\end{equation}
Therefore, at equilibrium, when the floater is above the rotor, we have $\theta>\gamma $; when it is below the rotor, we have $\theta<\gamma$.

\begin{figure}[tb]
    \centering
    \includegraphics[scale=0.55]{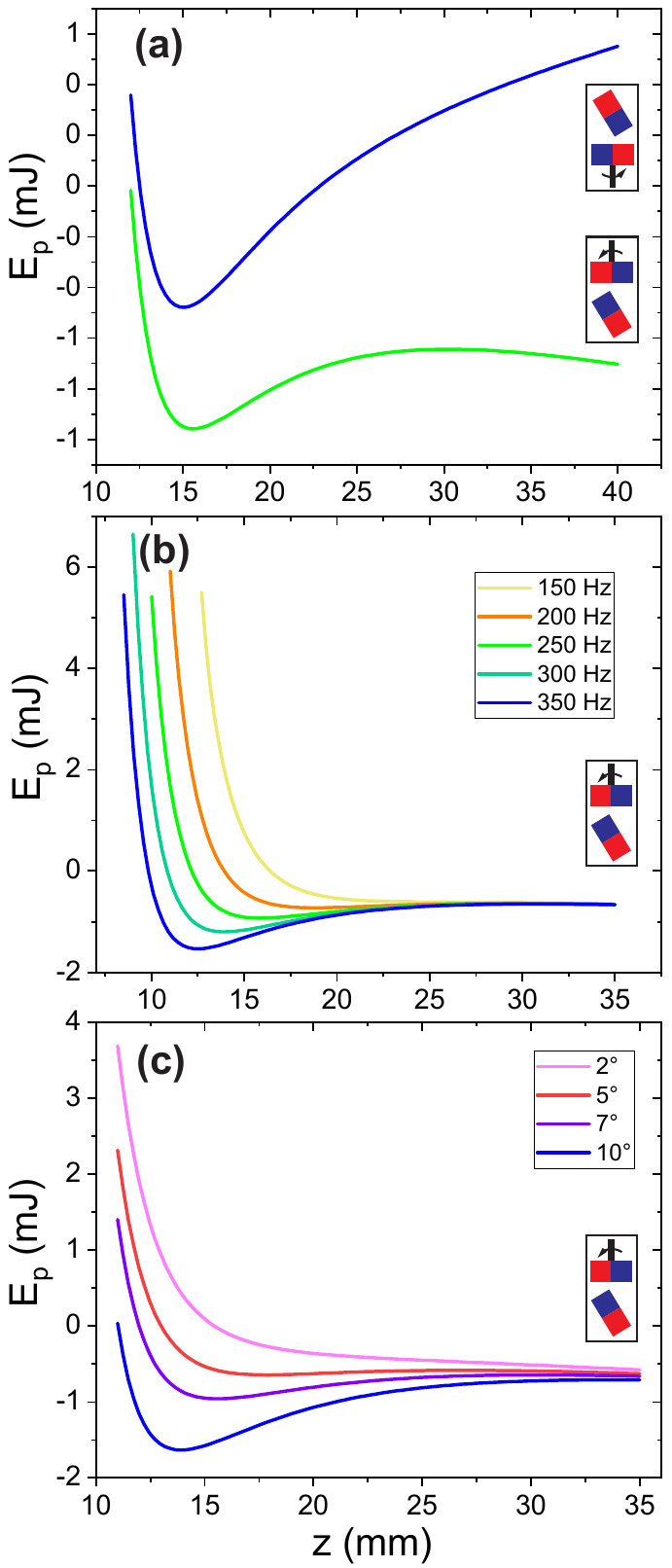}  
    \caption{Potential energy $E_p$ from Eq.~\eqref{genformula2} as a function of the distance $z$ between the two magnets along the $z$-axis, for a 6\,mm-side cubic floater magnet ($\mu_r = 0,955$ A$\cdot$ m$^2$, $\mu_f = 0,206$ A$\cdot$m$^2$, $m = 1.6\,$g). \textbf{(a)} The floater levitates over the rotor (blue), and below the rotor (green) ($f = 250$\,Hz, $\gamma = 7^\circ$). \textbf{(b)} The floater is below the rotor at different rotational frequencies for $\gamma = 7^\circ$ (yellow: $f=150$ Hz; orange: $f= 200$\,Hz; light green: $f=250$\,Hz; green: $f=300$\,Hz; blue: $f= 350$\,Hz). \textbf{(c)} The floater is below the rotor for different angles $\gamma$ at $f = 250$\,Hz (pink: $\gamma = 2^{\circ}$; red: $\gamma = 5^{\circ}$; purple: $\gamma = 7^{\circ}$; dark blue: $\gamma = 10^{\circ}$).}
    \label{energyz}
\end{figure} 

Fig.~\ref{energyz} shows the evolution of $ E_p (r=0,z)$ for different rotor/floater configurations, using Eq.\eqref{genformula2}. Fig.~\ref{energyz}a highlights that the most stable configuration is when the floater is above the rotor, because gravity helps to create a deeper potential well. Fig.~\ref{energyz}b shows the existence of a minimum rotational speed to obtain a potential well and thus a stable equilibrium around 200\,Hz in the present simulation. The higher the rotational speed, the deeper and narrower the potential well and the more stable the bounded state. Finally, Fig.~\ref{energyz}c emphasizes the need for an angle $\gamma$ large enough to create a potential well that would keep the floater in a bound state.

\subsection{Lateral Equilibrium close to the axis of rotation }\label{lateq}

When $r>0$, the solution $\alpha = \omega t $ and $\theta$ constant do not solve Eq. \eqref{lagalpha} and \eqref{lagphi} anymore.  Hermansen \textit{et al}  \cite{hermansen2023magnetic} suggested to add a constant dephasing angle $\psi$ such as $\alpha = \omega t + \psi$. In our model, this solution still fails to solve the motion equations \eqref{lagphi2} and \eqref{lagphi3}.

However, we chose to do the approximation of keeping the solution $\alpha(r,z,t) \approx \omega t$ and $\theta(r,z,t) = \theta_0(z)$ in Eq. \eqref{phi}. Thus, the potential energy in Eq. \eqref{genformula1} becomes
\begin{align}
E_p & =  -\frac{\mu_0 \mu_r \mu_f}{4\pi(r^2+z^2)^{\frac{5}{2}}} \Big\{  z^2 \big[ 2\cos\theta_0 \sin\gamma - \sin\theta_0 \cos\gamma \big] \nonumber \\
& \quad + 3rz \big[ \sin \gamma \sin\theta_0 + \cos\theta_0\cos\gamma \big] \cos(\omega t )  \nonumber \\
& \quad +  r^2 \big[  \cos\gamma \sin\theta_0 \; [2 \cos^2(\omega t) -  \sin^2(\omega t)] - \sin\gamma \cos\theta_0 \big] \Big\} \pm mgz. 
\end{align}

Experimentally, it can be observed that the center of mass of the floater moves slowly compared to the rotation of the rotor. The frequency of these small oscillations rarely exceeds 20\,Hz, while the rotor rotates hundreds of times per second. Therefore, the center of mass is not affected at first order by the rapid changes of the magnetic potential energy, but by its average over one rotation of the rotor, which leads to
\begin{align}
\langle E_p \rangle & =  -\frac{\mu_0 \mu_r \mu_f}{4\pi(r^2+z^2)^{\frac{5}{2}}}  \big(z^2 - \frac{1}{2}r^2 \big) \big[ 2\cos\theta_0 \sin\gamma - \sin\theta_0 \cos\gamma \big] \pm mgz. \label{avEn}
\end{align}
In Eq.~\eqref{avEn}, the energy is no longer time-dependent. The problem has been reduced to two spatial coordinates with cylindrical symmetry. As shown in Appendix B, the lateral equilibrium for $r = 0 $ is still ensured by the same condition as in Eq. \eqref{condition1}. Both the lateral equilibrium and the horizontal equilibrium discussed in Section II.B are stable. Thus, the Hessian matrix is positive definite at $(z_{eq} , 0 ) $ and the equilibrium is stable. It is important to note that $ \dfrac{d}{dz}\dfrac{d}{dr} \langle E_p \rangle (r=0,z_{eq}) = 0 $.

\subsection{Out of equilibrium movement of the floater center of mass} \label{offeq}

When externally perturbed, the floater can oscillate around its lateral or vertical equilibrium positions. As mentioned in \cite{hermansen2023magnetic}, the floater can have different types of motion. This section brings new insight into the behavior of the floater's center of mass away from the axis of rotation. It is experimentally observed that for small perturbations, the floater can oscillate around the axis of rotation remaining in a bounded state with the rotor and eventually returning to its equilibrium position if the damping of the medium is large enough. For larger perturbations, the floater is ejected.

In the following, the spatial limitation of this bounded state is studied.

\subsubsection{Vertical oscillations along the rotation axis. }

First, in order for the levitation to be possible as described in Section \ref{conical} the height should respect Eq. \eqref{omega0}. For a given angular velocity $\omega$, the minimal distance $z$ of equilibrium is then 
\begin{equation}
z_{min}^3 = \frac{\mu_0 \mu_r \mu_f}{8\pi I \omega^2} \cos \gamma \Big[ 1 + (2\tan \gamma)^{2/3} \Big]^{3/2} . \label{zmin}
\end{equation}

\begin{figure}[tb]
    \centering
 \includegraphics[scale=0.6]{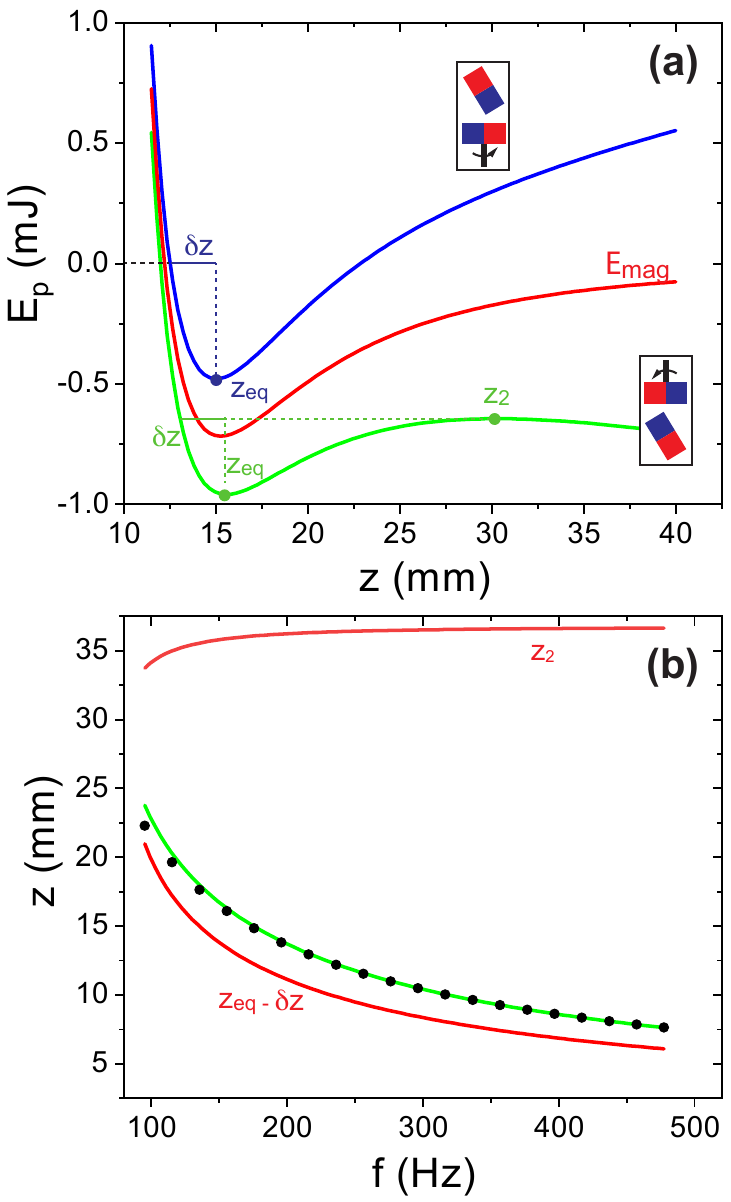} 
    \caption{\textbf{(a)} Potential energy  $ E_p$ from Eq.~\eqref{E_AB} as a function of the distance $z$ between the two magnets along the $z$-axis at $f = 250$\,Hz. The floater levitates above the rotor (blue), below the rotor (green), or the gravity is neglected (red). \textbf{(b)} Evolution of $z_{eq}$ (green), and the interval of stability $[z_{eq} - \delta z, z_2]$ (red) with respect to $f$. The black dots represent the approximation in Eq.~\eqref{zeq}. All the data are calculated for a 6\,mm-side cubic floater magnet ($\mu_r = 0,955$ A$\cdot$ m$^2$, $\mu_f = 0,206$ A$\cdot$m$^2$, $m = 1.6\,$g, $\gamma = 7^\circ$).} \label{delta_z} 
\end{figure}

At first-order approximation $\theta \ll 1 $, Eq.~\eqref{smallangle} rewrites as
\begin{equation}\label{F_AB}
 F_z  (r=0,z) = \mp mg - \frac{A}{z^4} + \frac{B(\omega)}{z^7}, 
\end{equation}
so the averaged potential energy is
\begin{equation}\label{E_AB}
 E_p  (r=0,z) = \pm mgz - \frac{A}{3z^3} + \frac{B(\omega)}{6z^6}, 
\end{equation}
with $A=\dfrac{ 3 \mu_0 \mu_r \mu_f \sin\gamma}{2\pi} >0$ and $B(\omega)=\dfrac{ 3\left( \mu_0 \mu_r \mu_f \cos\gamma \right)^2 }{ 16 \pi^2 I \omega^2} > 0$ decreasing with $\omega$ with $\lim_{\omega \rightarrow \infty} B(\omega)= 0$. Figure~\ref{delta_z}a shows the potential energy $ E_p$ for 3 cases: the floater above (blue line) or below (green line) the rotor, and gravity is neglected (red line).

When the floater is below the rotor, if $\omega > \omega_2$, levitation is possible but gravity can eject the floater vertically if its potential energy exceeds the local maximum. We note this position $z_2$ (see Fig.~\ref{delta_z}a). Thus, in this case, $ F_z (r=0,z) = 0 $ has two solutions $(z_{eq}, z_2)$ with $z_{eq} < z_2$. Assuming $\dfrac{B(\omega)}{z_2^7} \ll mg $, then $z_2 \simeq \left( \frac{A}{mg} \right)^{1/4}$ is independent of $\omega$ while $z_{eq}$ decreases with $\omega$ (see Fig.~\ref{delta_z}b). Thus, $z_2 - z_{eq}$ increases with $\omega$ (see also Fig.~\ref{delta_z}b).
The solution $z_{2}$ is the maximum distance for the two magnets to remain in the bound state. Hence, the higher the rotational speed, the larger is the interval of stability $[z_{eq}, z_2]$.

We can add more details on the stability depending on the position of the floater. When the floater is below the rotor, since $\lim_{z \rightarrow 0^+}  E_p = + \infty $, there exists $\delta z  > 0 $ such that $ E_p  (z_{eq} - \delta z ) =  E_p  (z_{2}  )  $ (see Fig.~\ref{delta_z}a, green line). Also since $\lim_{ \omega \rightarrow  + \infty} z_{eq}(\omega) = 0 $ and $z_{eq} - \delta z \in (0, z_{eq}) $, we can conclude that $\lim_{ \omega \rightarrow +\infty } \delta z (\omega) = 0 $. Therefore, the interval of stability $[z_{eq} - \delta z , z_{eq}]$ decreases towards 0 with increasing $\omega$. Therefore, at high speed if the floater reaches a position a little bit closer to the rotor than the equilibrium distance, it risks to be ejected. Fig.~\ref{delta_z}b confirms  that from a certain rotational speed $z_2$ is independent of $\omega $ and highlights the fact that neglecting the weight around $z_{eq}$ can be an efficient approximation. 

If the floater is above the rotor, $\lim_{z \rightarrow + \infty}  E_p = +\infty$ and $\lim_{z \rightarrow 0^+}  E_p = +\infty$ (see Fig.~\ref{delta_z}a, blue line). Hence, the floater should never be ejected vertically. However, the oscillation is experimentally never vertical only and small lateral perturbations are enhanced when the floater is the furthest away from the rotor. This can ultimately result in lateral ejection. As shown in Fig.~\ref{delta_z}a, it is also possible to define $\delta z $ when the floater is above the rotor by considering the floater to be out of the bound state when $ F_z  \approx mg$. In the convention used so far, $\lim_{z \rightarrow \infty}  E_{mag} = 0$. Therefore, $\delta z$ is defined as the positive solution, such that $ E_{mag}  (r=0, z_{eq} - \delta z ) = 0$. 

We conclude that, at high rotational speed, as shown in Fig. \ref{energyz}, the floater is more stable but cannot move as much. At lower rotational speed the floater can move more around its equilibrium position as long as it respects Eq.\,\eqref{zmin}.

\subsubsection{Lateral oscillations}

We have just established that, when the floater is above the rotor, it cannot be ejected vertically. Experimentally, when the perturbation is large enough, the bounding may collapse. Therefore, it is necessary to study the equilibrium out of the rotational axis. From Eq. \eqref{avEn} we can deduce the lateral force
\begin{align}
\langle F_r \rangle 
&=
\frac{3\,\mu_0\,\mu_r\,\mu_f}{8\pi}\;
\frac{
\,(4z^2 - r^2)\;
\big(\sin\theta_0 \cos\gamma - 2\cos\theta_0 \sin\gamma \big)
}{
(r^2 + z^2)^{7/2}
}r. \label{fr_av}
\end{align}

This equation shows an attractive force up until $r = 2z $ and then a repelling force. Experimentally, the floater does not always reach $r = 2z $, and falls at smaller $r$. To explain this phenomenon, we must study the vertical force. Using Eqs.\eqref{phi} and \eqref{genformula2}, we derive
\begin{align}
\langle F_z \rangle = & I \omega^2 \, \frac{z^2 \sin \theta_0 \cos \theta_0}{(r^2 + z^2)^{7/2}}
\Bigg[
\, z^2(6z^2 - 9r^2 ) \,
\frac{(\cos\gamma \sin\theta_0 - 2\cos\theta_0 \sin\gamma)}
{\cos\gamma \cos\theta_0 + 2\sin\gamma \sin\theta_0}
\nonumber \\ &+ 6 (r^2 + z^2)(z^2 - \tfrac{1}{2}r^2 )\,
\frac{(\cos\gamma \cos\theta_0 + 2\sin\gamma \sin\theta_0) \sin \theta_0 \cos \theta_0\,}
{ \cos\gamma \cos^3\theta_0 - 2\sin\gamma \sin^3 \theta_0 }
\Bigg]
\mp mg. \label{Fzofr}
\end{align}

\begin{figure}[tb]
    \centering    \includegraphics[scale=1]{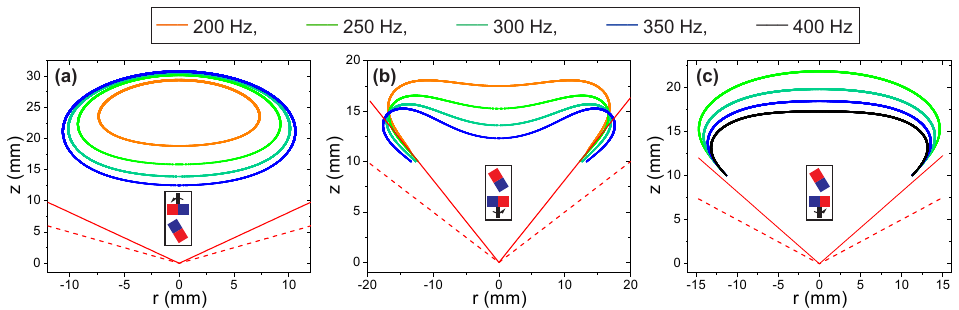} 
    \caption{  Solution of the equation $\langle F_z \rangle = 0 $ from Eq.\,\ref{Fzofr}, in the $(r,z)$ plane for various rotational speed (orange: $f=200$\,Hz; light green: $f=250$\,Hz; green: $f=300$; blue: $f= 350$\,Hz, and black: $f=400$\,Hz). \textbf{(a)} and \textbf{(b)} for a 6\,mm-side cubic floater magnet ($\mu_r = 0,955$ A$\cdot$ m$^2$, $\mu_f = 0,206$ A$\cdot$m$^2$, $m = 1.6\,$g, $\gamma = 7^{\circ}$) and \textbf{(c)} for a 5\,mm-side cubic floater magnet ($\mu_r = 0,955$ A$\cdot$ m$^2$, $\mu_f = 0,120$ A$\cdot$m$^2$, $m = 0.95\,$g, $\gamma = 2^{\circ}$). The floater levitates below the rotor \textbf{(a)} , or above the rotor \textbf{(b)} and \textbf{(c)}.  The solid red line represents the equation $r = \sqrt{3/2}z $ and the dotted red line the equation $r = 2z$.}
    \label{corazon}
\end{figure} 

In this equation we see that, under the condition of Eq.\eqref{condition1}, if the floater is below the rotor, there is no stable point for $r$ exceeding $\sqrt{3/2} z$. For $\langle F_z \rangle = 0 $, increasing the rotational speed has the same effect as decreasing the gravitational force.

\begin{figure}[tb]
    \centering    \includegraphics[scale=0.6]{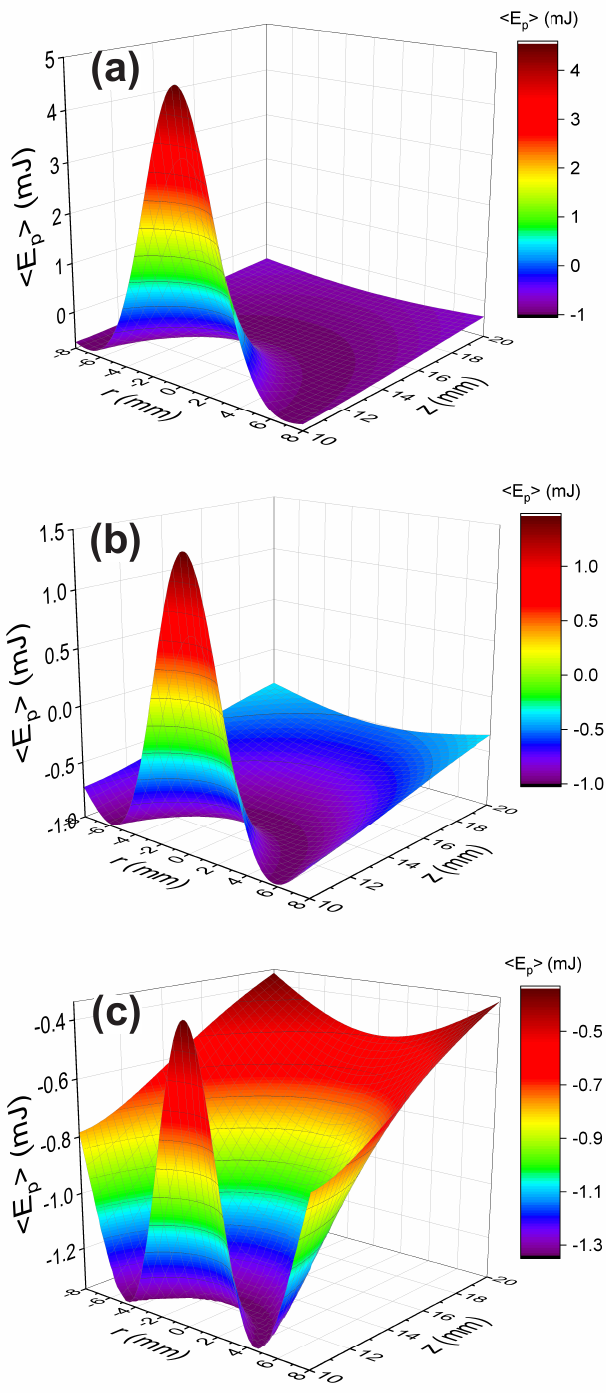} 
    \caption{Average energy $\langle E_p \rangle$ from Eq.~\eqref{avEn} as a function of $r$ and $z$ for a 6\,mm-side cubic floater magnet for different rotor frequencies ($\mu_r = 0,955$ A$\cdot$m$^2$, $\mu_f = 0,206   $ A$\cdot$m$^2$, $\gamma = 7^\circ$ and $m = 1.6\,g $). \textbf{(a)} $f = 200 $ Hz, \textbf{(b)} $f =250$ Hz and \textbf{(c)} $f = 350$ Hz.}
    \label{3D}
\end{figure} 

Fig.~\ref{corazon} shows the solutions to the equation $\langle F_z \rangle = 0 $ for different floaters at various rotational speeds, both above and below the rotor. When the floater is below the rotor, the solutions to the equation $\langle F_z \rangle = 0 $ get closer to the limit $r_{max} = \sqrt{3/2} z $ as the rotational speed increases. However, when the floater is above the rotor, this limit can be exceeded (see Fig.~\ref{corazon}b). We observe that the curve of the equation $\langle F_z \rangle = 0 $, $r_{max}$ can increase (Fig.~\ref{corazon}b) or decrease (Fig.~\ref{corazon}c) with the rotation speed. 


Fig.~\ref{3D} confirms that increasing the rotational speed deepens the potential well. Therefore, at low velocities, the floater is more mobile and the levitation is less stable. At higher velocities, the well becomes U-shaped. This qualitatively corresponds to the experimental results of Hermansen \textit{et al} \cite{hermansen2023magnetic}.

\section{Experimental results}

\begin{figure}[tbh]
    \centering
    \includegraphics[scale=0.33]{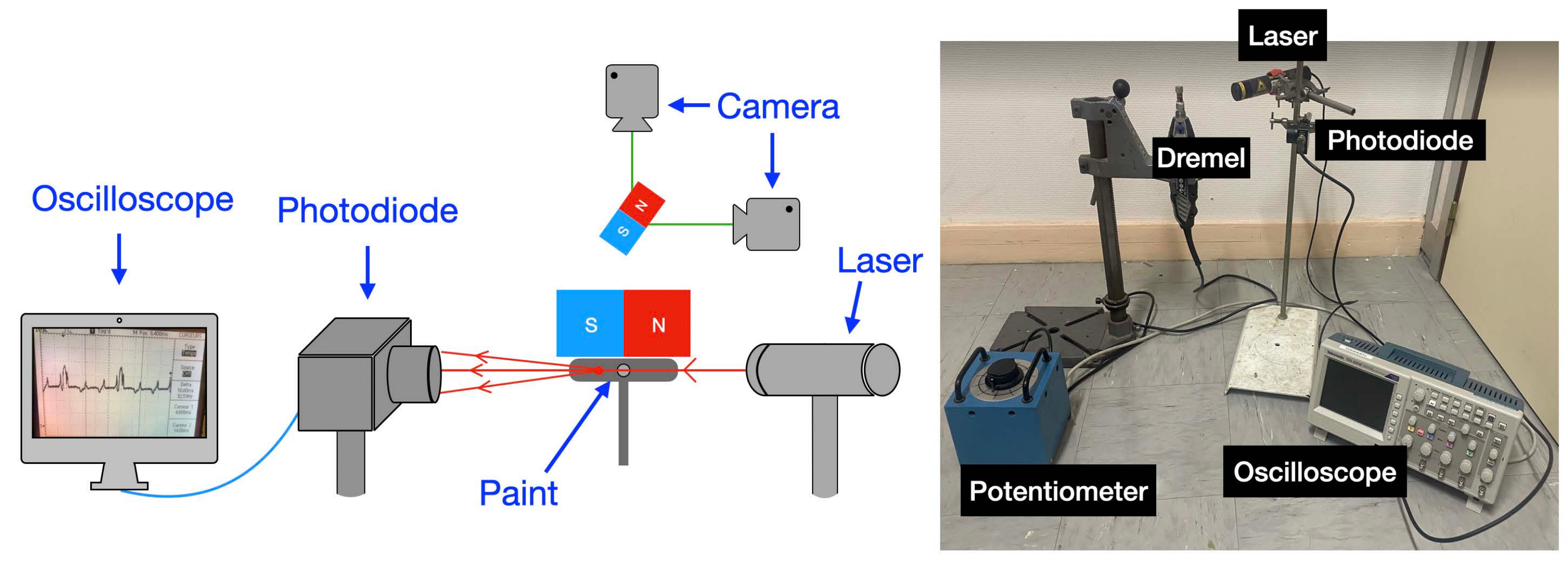} 
    \caption{ Scheme of the experimental set-up for the measurement of the rotational frequency on the left and a picture of it on the right.}
    \label{setup}
\end{figure}

To experimentally investigate the phenomenon, a 10 mm side cubic neodymium NdFeB permanent magnet rotor (Supermagnet, N42) with a remanence of 1.29 - 1.32 T was mounted on a high-speed die grinder (DREMEL 4200). Several permanent magnets of the same remanence were used for the floaters. One side of the rotating part of the high-speed grinder was painted white to measure its rotational frequency using a HeNe laser and a photodiode (Thorlabs PDA10A2) connected to an oscilloscope, as shown in Fig.~\ref{setup}. Three cameras were used during the experiments: a high-speed camera (3000 frames per second) to study the conical trajectory of the floater \cite{slowmo_hugo}, and two smartphone cameras (iPhones 12, 30 frames per second), one placed vertically and the other horizontally, to monitor the movement of the floater magnet.  A bubble level ($\pm2^\circ $) ensures that the axis of rotation is vertical.

\subsection{ Levitation spans} 
\begin{figure}[tbh]
   \centering
  \includegraphics[scale=1]{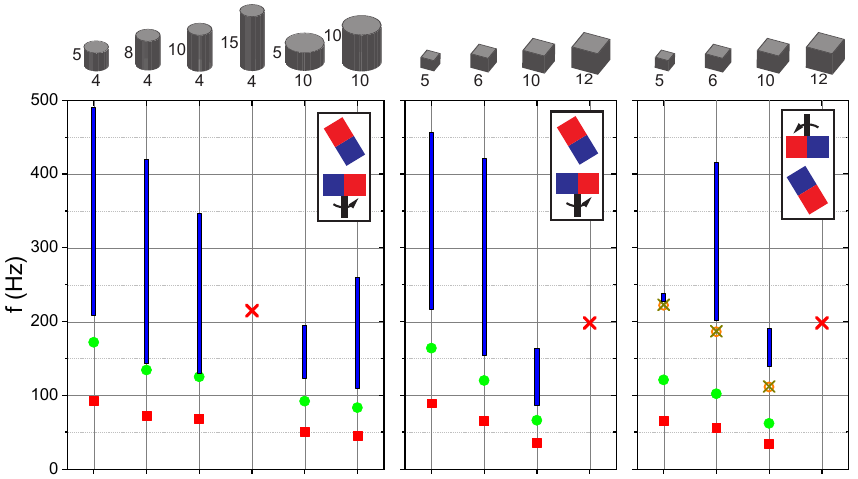} 
   \caption{Magnet levitation domains of cylindrical and cubic floaters (size is in mm). In the left and center figures, the floater is above the rotor. In the right figure, the floater is below the rotor. The blue bars are experimental stability ranges. The red crosses indicate that no levitation was observed. Red squares are theoretical values of $f_0 = \omega_0/2\pi$, from Eq.~\eqref{omega0}, green rounds are theoretical values of $f_{1} = \omega_{1}/2\pi $, from Eq.~\eqref{omega1}, orange circles are the theoretical values of $f_2' = \omega_{2}'/2\pi$, from Eq.~\eqref{omega2prime}, and dark green crosses are the computed values of $f_2 = \omega_{2}/2\pi$ the first frequency which allows $F_z (r= 0, z) = 0$ to have a solution in Eq.~\eqref{smallangle}. All these theoretical values are calculated using $\gamma = 5.5^\circ$ and $z_{eq}$ the theoretical equilibrium distance calculated using the rotational frequency of the first observed levitation.}
    \label{span}
\end{figure}

Levitation domains have been studied for several shapes and sizes of floater magnets, depending on the rotational speed. Once the die grinder has reached a constant velocity, the floater is brought close to the rotor using a plastic plate. Levitation is considered possible if, out of five tests, at least two one-second levitation have been observed. Multiple measurements were taken every approximately 10\,Hz. The results are shown in Fig.~\ref{span}. 

As predicted by theory, stability domains are wider when the floater is above the rotor. Also, the larger the magnet, the lower the rotational speed required for levitation. This is to be expected from the dependence on $I^{-1/2}$ of $\omega_0$ , $\omega_{1}$ and $\omega_2'$ in Eqs.~\eqref{omega0},  \eqref{omega1} and \eqref{omega2prime}. The lower limit $f_0$ is significantly smaller than the first levitation observation. This was also to be expected, as it corresponds to an angle $\theta$ close to $90^\circ $, but $f_1$ is a more precise lower bound. When the floater is below the rotor, faster rotation is required to compensate for gravity, and this new constraint corresponds to a new lower bound $\omega_2$ which can be well approximated by $\omega_2'$.

\begin{figure}[tbh]
   \centering
  \includegraphics[scale=0.8]{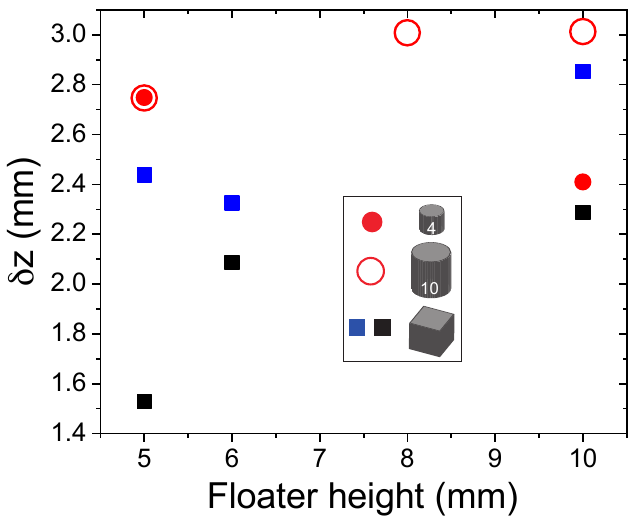} 
   \caption{Theoretical value of $\delta z$ taken for the maximum rotational frequency where levitation was observed for each magnet in Fig.~\ref{span}, as a function of the size of the height of the floater, for the 4\,mm large cylinders above the rotor (small red rounds), the 10\,mm large cylinders above the rotor (open red rounds), the 10\,mm cube above the rotor (blue squares) and the cube below the rotor (black squares). The two theoretical values of $ \delta z $ for the cylinders of height 5\,mm are almost identical.}
    \label{delta_zzz}
\end{figure}

The high-speed motor used is able to reach 550 Hz, nevertheless none of the magnets tested can levitate at velocity higher than 500 Hz. Each one has an upper boundary. One can observe that the bigger the floater is, the smaller its upper boundary is. As predicted in section~\ref{offeq}, the potential well becomes too narrow, the area of stable levitation too small and a bigger floater gets ejected more easily. This is visible in Fig.~\ref{delta_zzz}. This figure shows that the smaller the magnet, the smaller $\delta z$ can be before the magnet is ejected. This explains the upper rotational speed limit for levitation phenomena. When the floater is below the rotor, gravity helps eject it. As a result, the area of stable levitation and $\delta z $ are smaller. For a 10 mm cubic floater, the upper bound is smaller when the floater is above the rotor. This is because $z_{eq}$ is smaller when the floater is above the rotor.

\subsection{Motion of the floater}

\begin{figure}[tbh]
    \centering
    \includegraphics[scale=0.6]{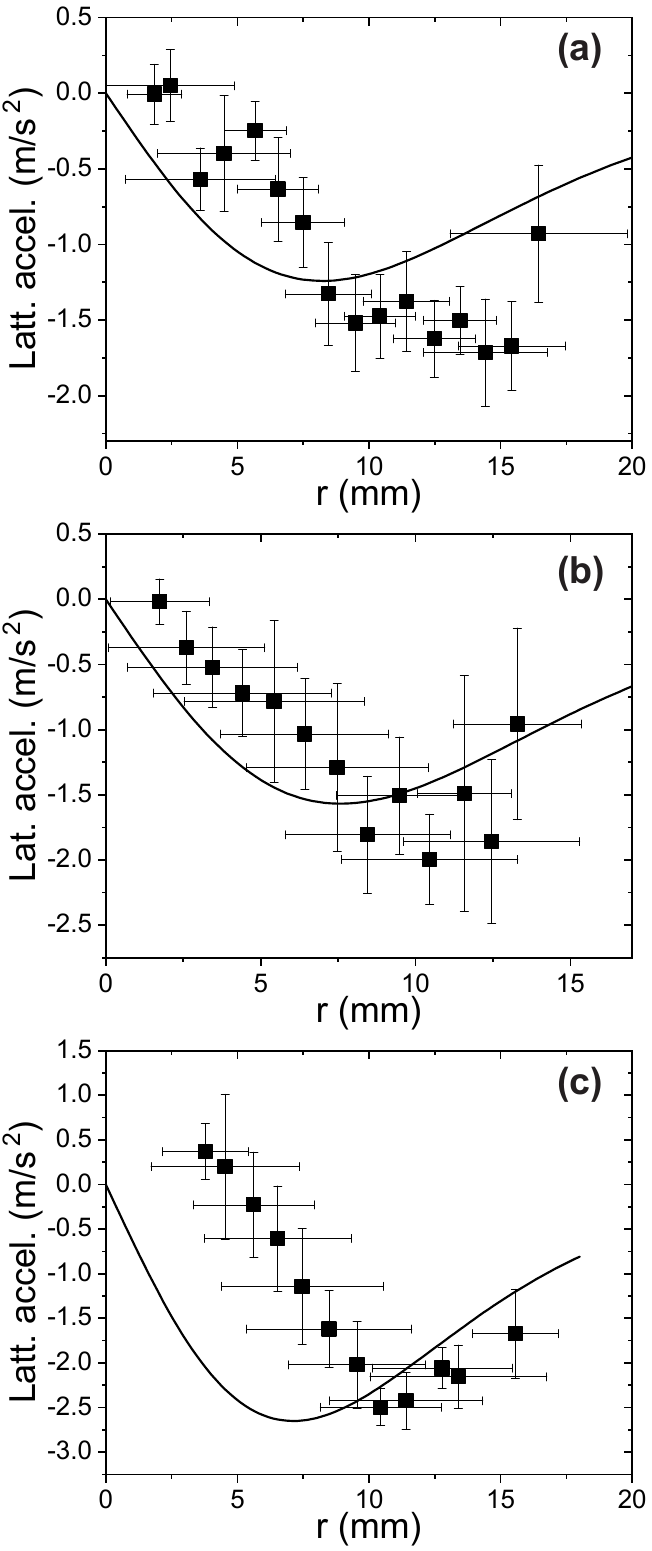} 
    \caption{Measurements of the mean lateral acceleration as a function of distance $r$ from the $z$ axis of a 
5\,mm cubic floater. Experiments (squares), and theory (solid lines) for $\langle F_r \rangle/m$ using Eq.~\eqref{fr_av}. Rotational frequencies are \textbf{(a)} $f = 238\pm5$\,Hz ($\gamma = 2.4^{\circ}$); \textbf{(b)} $f = 307\pm2$\,Hz ($\gamma = 1.9^{\circ}$) and \textbf{(c)} $f = 351\pm2$\,Hz ($\gamma = 1.9^{\circ}$).}
    \label{leftright}
\end{figure}

In this section, the motion of a 5 mm cubic floater is studied out of equilibrium. Measurements are limited to three frequencies: $238\pm5$\,Hz, $307\pm2$\, Hz and $351\pm2$\,Hz.
The configuration may be slightly different in that, between this experiment and the previous one, the rotor magnet was peeled off and glued back on from the grinder bit. This may have had an effect on the value of $\gamma$.
In order to study lateral movements for each frequency, variations in floater height that are small in relation to its lateral variations were tracked by video. Here, $z = z_{eq}$ is taken as constant. Fig.~\ref{leftright} shows that the higher the rotational speed, the greater the lateral force. Thus, the higher the velocity, the better the floater is trapped which corresponds to the description given in section~\ref{offeq}.  We chose $\gamma$ for the best fit. At $f = 351\pm2$\,Hz, the theoretical values do not fit as well the experimental data. This result may be explained by the frequency of the oscillations getting too large compared to the sampling of the camera leading to measurement uncertainties.

\begin{figure}[tbh]
    \centering
    \includegraphics[scale=0.80]{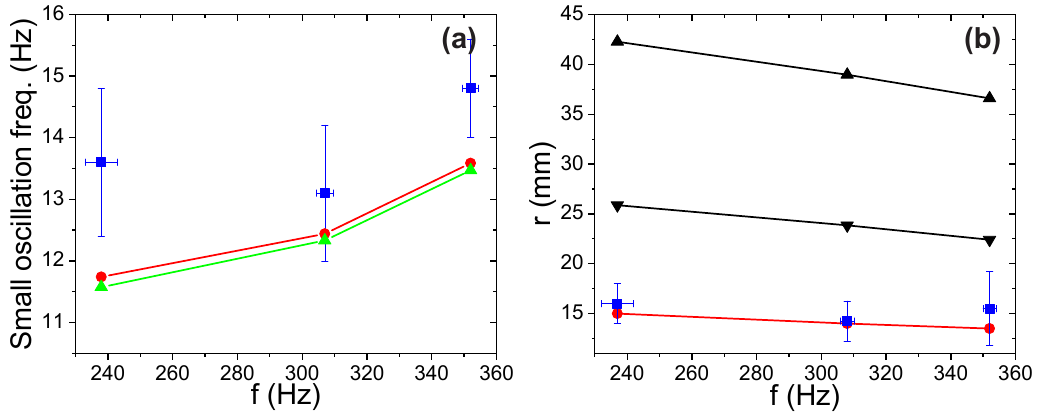} 
    \caption{\textbf{(a)} Experimental measures of the vertical small oscillations (blue squares), theoretical values using Eq.\,\eqref{Fz_exact} (red rounds), and theoretical values for small angle approximation Eq.\,\eqref{smallangle} (green triangles). \textbf{(b)} Maximum value of $r$ measured during levitation (blue squares), theoretical $r_{max}$ using Eq. \eqref{Fzofr} (red rounds), equation $r = \sqrt{3/2} z_{eq}$ (black down triangles), and equation $r = 2z_{eq}$ (black up triangles). The same angles $\gamma $ have been used as Fig. \ref{leftright}.}
    \label{more_res}
\end{figure}

Other parameters can be extracted from the experimental data. For each of the three frequencies, the vertical oscillations frequency has been measured for 10 different videos and compared to the theoretical ones that can be extracted from Eq. \eqref{Fz_exact} and equation from Eq. \eqref{smallangle} in the small angle approximation. The results are displayed in Fig. \ref{more_res}a. Similarly as in Fig. \ref{delta_z}, the small angle approximation gives very similar results.

Figure \ref{more_res}b compares the maximum experimental values of $r$ with the limits $r=2z_{eq}$ and $r=\sqrt{3/2}z_{eq}$ and the theoretical value of $r_{max}$ obtained from Eq.~\eqref{Fzofr}. As predicted in Section~\ref{offeq}, the first two limits are poor approximations of the experimental data, while the third limit is much better. Note that levitation is observed beyond the last limit. This could be due to the approximation $\theta \approx \theta_0$ no longer holding for greater values of $r$, or to the rotor's squared shape not exactly following the dipole approximation.

\section{Conclusion}

We have shown that under certain conditions of size and rotational speed, a permanent magnet can levitate above and below another rotating permanent magnet. In particular, we have refined the lower limit below which levitation is not possible, and thus obtained 3 theoretical values with increasing precision. We have extended the theory of the out of equilibrium motion of the floater, and described the upper angular velocity limit of stability. We experimentally investigated the levitation range of rotational speed for several floater's sizes finding a good agreement with our theoretical results for the out of equilibrium motion of the floater. Further investigations including the role and control of the angle $\gamma$ and the study of off-axis floater orientation would need to be done to understand fully this phenomenon. 

\section{Appendixes}

\subsection{Calculation of Eq. \eqref{omega0}}
We isolate \( \omega^{2} \) in  Eq.\eqref{phi}:

\begin{equation}
    \omega^{2} = A \left[ \frac{2 \sin (\gamma)}{\cos (\theta)} + \frac{\cos \gamma}{ \sin (\theta)} \right],
    \quad \text{with} \quad
    A = \frac{\mu_{0} \mu_{n} \mu_{1}}{2 I (4 \pi z^3)}
\end{equation}

We consider \( \omega^{2} \) as a function of \( \theta \) and find for which \( \theta \) it is minimized:

\begin{equation}
    \frac{\partial }{\partial \theta} \omega^{2} = A \left( \frac{2 \sin (\gamma) \sin (\theta)}{\cos^{2}(\theta)} - \frac{\cos \gamma \cos \theta}{\sin^{2}(\theta)} \right)
\end{equation}

\( \omega^{2} \) reaches a minimum for:

\begin{equation}
    \theta_{\text{min}} = \text{Arctan} \left( (2 \tan(\gamma))^{-1/3} \right)
\end{equation}

 We note that when $\gamma \ll 1$ we have $\theta_{\min} \simeq 90^{\circ}$. We note : 
\[
t=\tan\gamma,\qquad s=\tan\theta_{\min}=(2t)^{-1/3}.
\]
Using the standard relations
\[
\sin\gamma=\frac{t}{\sqrt{1+t^{2}}},\quad
\cos\gamma=\frac{1}{\sqrt{1+t^{2}}},\quad
\sin\theta=\frac{s}{\sqrt{1+s^{2}}},\quad
\cos\theta=\frac{1}{\sqrt{1+s^{2}}}.
\]
Then the bracket becomes
\begin{align*}
\frac{2\sin\gamma}{\cos\theta}+\frac{\cos\gamma}{\sin\theta}
&= \frac{2\sin\gamma}{1/\sqrt{1+s^{2}}} + \frac{\cos\gamma}{s/\sqrt{1+s^{2}}}\\[4pt]
&= \frac{\sqrt{1+s^{2}}}{\sqrt{1+t^{2}}}\!\left(2t+\frac{1}{s}\right).
\end{align*}

Now substitute \(s=(2t)^{-1/3}\). Let
\[
u=(2t)^{1/3},\qquad \text{so } s=\frac{1}{u},\quad 2t=u^{3}.
\]
Then
\[
2t+\frac{1}{s}=u^{3}+u = u(1+u^{2}),
\qquad
1+s^{2}=1+\frac{1}{u^{2}}=\frac{1+u^{2}}{u^{2}}.
\]
Thus
\[
\frac{\sqrt{1+s^{2}}}{\sqrt{1+t^{2}}}\Big(2t+\frac{1}{s}\Big)
= \frac{\sqrt{1+u^{2}}}{u\sqrt{1+t^{2}}}\cdot u(1+u^{2})
= \frac{(1+u^{2})^{3/2}}{\sqrt{1+t^{2}}}.
\]

Undo the definition \(u=(2t)^{1/3}\):
\[
1+u^{2}=1+(2t)^{2/3}=1+(2\tan\gamma)^{2/3},
\qquad
\sqrt{1+t^{2}}=\sqrt{1+\tan^{2}\gamma}=\frac{1}{|\cos\gamma|}.
\]

Hence we obtain the compact formula
\[
\omega^{2}
= A\,|\cos\gamma|\;\bigl(1+(2\tan\gamma)^{2/3}\bigr)^{3/2}.
\]

\subsection{Proof that $\tan \theta \leq 2 \tan \gamma$ is a condition of lateral stable equilibrium for Eq. \eqref{avEn}}

We expend $\langle E_p \rangle$ to the second order in $r \ll z $ : 

\begin{align}
\langle E_p \rangle & =  -\frac{\mu_0 \mu_r \mu_f}{4\pi(r^2+z^2)^{\frac{5}{2}}}  \big(z^2 - \frac{1}{2}r^2 \big) \big[ 2\cos\theta_0 \sin\gamma - \sin\theta_0 \cos\gamma \big] \pm mgz \nonumber \\ 
 & \approx  -\frac{\mu_0 \mu_r \mu_f}{4\pi z^5} \big( 1 - \frac{5}{2}\frac{r^2}{z^2}  \big)  \big(z^2 - \frac{1}{2}r^2 \big) \big[ 2\cos\theta_0 \sin\gamma - \sin\theta_0 \cos\gamma \big] \pm mgz \nonumber \\  
 & \approx  -\frac{\mu_0 \mu_r \mu_f}{4\pi z^3}  \big[ 2\cos\theta_0 \sin\gamma - \sin\theta_0 \cos\gamma \big] \pm mgz + r^2 \frac{3\mu_0 \mu_r \mu_f}{4\pi z^5}    \big[ 2\cos\theta_0 \sin\gamma - \sin\theta_0 \cos\gamma \big] \nonumber
\end{align}

Thus there is a stable lateral equilibrium if and only if $ 2\cos\theta_0 \sin\gamma - \sin\theta_0 \cos\gamma > 0 $ if and only if $\tan \theta \leq 2 \tan \gamma$. 

\hfill $\square$

\bibliography{biblio}

\begin{thebibliography}{20}%
\makeatletter
\providecommand \@ifxundefined [1]{%
 \@ifx{#1\undefined}
}%
\providecommand \@ifnum [1]{%
 \ifnum #1\expandafter \@firstoftwo
 \else \expandafter \@secondoftwo
 \fi
}%
\providecommand \@ifx [1]{%
 \ifx #1\expandafter \@firstoftwo
 \else \expandafter \@secondoftwo
 \fi
}%
\providecommand \natexlab [1]{#1}%
\providecommand \enquote  [1]{``#1''}%
\providecommand \bibnamefont  [1]{#1}%
\providecommand \bibfnamefont [1]{#1}%
\providecommand \citenamefont [1]{#1}%
\providecommand \href@noop [0]{\@secondoftwo}%
\providecommand \href [0]{\begingroup \@sanitize@url \@href}%
\providecommand \@href[1]{\@@startlink{#1}\@@href}%
\providecommand \@@href[1]{\endgroup#1\@@endlink}%
\providecommand \@sanitize@url [0]{\catcode `\\12\catcode `\$12\catcode `\&12\catcode `\#12\catcode `\^12\catcode `\_12\catcode `\%12\relax}%
\providecommand \@@startlink[1]{}%
\providecommand \@@endlink[0]{}%
\providecommand \url  [0]{\begingroup\@sanitize@url \@url }%
\providecommand \@url [1]{\endgroup\@href {#1}{\urlprefix }}%
\providecommand \urlprefix  [0]{URL }%
\providecommand \Eprint [0]{\href }%
\providecommand \doibase [0]{http://dx.doi.org/}%
\providecommand \selectlanguage [0]{\@gobble}%
\providecommand \bibinfo  [0]{\@secondoftwo}%
\providecommand \bibfield  [0]{\@secondoftwo}%
\providecommand \translation [1]{[#1]}%
\providecommand \BibitemOpen [0]{}%
\providecommand \bibitemStop [0]{}%
\providecommand \bibitemNoStop [0]{.\EOS\space}%
\providecommand \EOS [0]{\spacefactor3000\relax}%
\providecommand \BibitemShut  [1]{\csname bibitem#1\endcsname}%
\let\auto@bib@innerbib\@empty
\bibitem [{\citenamefont {Griffiths}(2023)}]{griffiths2023introduction}%
  \BibitemOpen
  \bibfield  {author} {\bibinfo {author} {\bibfnamefont {David~J}\ \bibnamefont {Griffiths}},\ }\href@noop {} {\emph {\bibinfo {title} {Introduction to electrodynamics}}}\ (\bibinfo  {publisher} {Cambridge University Press},\ \bibinfo {year} {2023})\BibitemShut {NoStop}%
\bibitem [{\citenamefont {Simon}\ and\ \citenamefont {Geim}(2000)}]{simon2000diamagnetic}%
  \BibitemOpen
  \bibfield  {author} {\bibinfo {author} {\bibfnamefont {Martin~D.}\ \bibnamefont {Simon}}\ and\ \bibinfo {author} {\bibfnamefont {Andre~K.}\ \bibnamefont {Geim}},\ }\bibfield  {title} {\enquote {\bibinfo {title} {Diamagnetic levitation: Flying frogs and floating magnets},}\ }\href@noop {} {\bibfield  {journal} {\bibinfo  {journal} {Journal of Applied Physics}\ }\textbf {\bibinfo {volume} {87}},\ \bibinfo {pages} {6200--6204} (\bibinfo {year} {2000})}\BibitemShut {NoStop}%
\bibitem [{\citenamefont {Kim}(2019)}]{kim2019levitation}%
  \BibitemOpen
  \bibfield  {author} {\bibinfo {author} {\bibfnamefont {Chan-Joong}\ \bibnamefont {Kim}},\ }\href@noop {} {\emph {\bibinfo {title} {Superconductor Levitation: Concepts and Experiments}}}\ (\bibinfo  {publisher} {Springer},\ \bibinfo {year} {2019})\ pp.\ \bibinfo {pages} {51--66}\BibitemShut {NoStop}%
\bibitem [{\citenamefont {Michaelis}\ \emph {et~al.}(2020)\citenamefont {Michaelis}, \citenamefont {Bingham}, \citenamefont {Charlton},\ and\ \citenamefont {Isaac}}]{michaelis2020variety}%
  \BibitemOpen
  \bibfield  {author} {\bibinfo {author} {\bibfnamefont {Max}\ \bibnamefont {Michaelis}}, \bibinfo {author} {\bibfnamefont {Bob}\ \bibnamefont {Bingham}}, \bibinfo {author} {\bibfnamefont {Mike}\ \bibnamefont {Charlton}}, \ and\ \bibinfo {author} {\bibfnamefont {C.~Aled}\ \bibnamefont {Isaac}},\ }\bibfield  {title} {\enquote {\bibinfo {title} {A variety of levitrons: a review},}\ }\href@noop {} {\bibfield  {journal} {\bibinfo  {journal} {European Journal of Physics}\ }\textbf {\bibinfo {volume} {42}},\ \bibinfo {pages} {015001} (\bibinfo {year} {2020})}\BibitemShut {NoStop}%
\bibitem [{\citenamefont {Simon}\ \emph {et~al.}(1997)\citenamefont {Simon}, \citenamefont {Heflinger},\ and\ \citenamefont {Ridgway}}]{simon1997spin}%
  \BibitemOpen
  \bibfield  {author} {\bibinfo {author} {\bibfnamefont {Martin~D.}\ \bibnamefont {Simon}}, \bibinfo {author} {\bibfnamefont {Lee~O.}\ \bibnamefont {Heflinger}}, \ and\ \bibinfo {author} {\bibfnamefont {S.L.}\ \bibnamefont {Ridgway}},\ }\bibfield  {title} {\enquote {\bibinfo {title} {Spin stabilized magnetic levitation},}\ }\href@noop {} {\bibfield  {journal} {\bibinfo  {journal} {American Journal of Physics}\ }\textbf {\bibinfo {volume} {65}},\ \bibinfo {pages} {286--292} (\bibinfo {year} {1997})}\BibitemShut {NoStop}%
\bibitem [{\citenamefont {Lee}\ \emph {et~al.}(2006)\citenamefont {Lee}, \citenamefont {Kim},\ and\ \citenamefont {Lee}}]{lee2006review}%
  \BibitemOpen
  \bibfield  {author} {\bibinfo {author} {\bibfnamefont {Hyung-Woo}\ \bibnamefont {Lee}}, \bibinfo {author} {\bibfnamefont {Ki-Chan}\ \bibnamefont {Kim}}, \ and\ \bibinfo {author} {\bibfnamefont {Ju}~\bibnamefont {Lee}},\ }\bibfield  {title} {\enquote {\bibinfo {title} {Review of maglev train technologies},}\ }\href@noop {} {\bibfield  {journal} {\bibinfo  {journal} {IEEE transactions on magnetics}\ }\textbf {\bibinfo {volume} {42}},\ \bibinfo {pages} {1917--1925} (\bibinfo {year} {2006})}\BibitemShut {NoStop}%
\bibitem [{\citenamefont {Schweitzer}\ \emph {et~al.}(2009)\citenamefont {Schweitzer}, \citenamefont {Maslen} \emph {et~al.}}]{schweitzer2009magnetic}%
  \BibitemOpen
  \bibfield  {author} {\bibinfo {author} {\bibfnamefont {Gerhard}\ \bibnamefont {Schweitzer}}, \bibinfo {author} {\bibfnamefont {Eric~H}\ \bibnamefont {Maslen}},  \emph {et~al.},\ }\href@noop {} {\emph {\bibinfo {title} {Magnetic bearings}}}\ (\bibinfo  {publisher} {Springer},\ \bibinfo {year} {2009})\BibitemShut {NoStop}%
\bibitem [{\citenamefont {Supreeth}\ \emph {et~al.}(2022)\citenamefont {Supreeth}, \citenamefont {Bekinal}, \citenamefont {Chandranna},\ and\ \citenamefont {Doddamani}}]{supreeth2022review}%
  \BibitemOpen
  \bibfield  {author} {\bibinfo {author} {\bibfnamefont {D.K.}\ \bibnamefont {Supreeth}}, \bibinfo {author} {\bibfnamefont {Siddappa~I.}\ \bibnamefont {Bekinal}}, \bibinfo {author} {\bibfnamefont {Shivamurthy~Rokkad}\ \bibnamefont {Chandranna}}, \ and\ \bibinfo {author} {\bibfnamefont {Mrityunjay}\ \bibnamefont {Doddamani}},\ }\bibfield  {title} {\enquote {\bibinfo {title} {A review of superconducting magnetic bearings and their application},}\ }\href@noop {} {\bibfield  {journal} {\bibinfo  {journal} {IEEE Transactions on Applied Superconductivity}\ }\textbf {\bibinfo {volume} {32}},\ \bibinfo {pages} {1--15} (\bibinfo {year} {2022})}\BibitemShut {NoStop}%
\bibitem [{\citenamefont {Shameli}(2008)}]{shameli2008design}%
  \BibitemOpen
  \bibfield  {author} {\bibinfo {author} {\bibfnamefont {Ehsan}\ \bibnamefont {Shameli}},\ }\emph {\bibinfo {title} {Design, implementation and control of a magnetic levitation device}},\ \href@noop {} {Ph.D. thesis},\ \bibinfo  {school} {University of Waterloo} (\bibinfo {year} {2008})\BibitemShut {NoStop}%
\bibitem [{\citenamefont {Wang}\ \emph {et~al.}(2019)\citenamefont {Wang}, \citenamefont {Lourette}, \citenamefont {O'Kelley}, \citenamefont {Kayci}, \citenamefont {Band}, \citenamefont {Kimball}, \citenamefont {Sushkov},\ and\ \citenamefont {Budker}}]{wangDynamicsFerromagneticParticle2019}%
  \BibitemOpen
  \bibfield  {author} {\bibinfo {author} {\bibfnamefont {Tao}\ \bibnamefont {Wang}}, \bibinfo {author} {\bibfnamefont {Sean}\ \bibnamefont {Lourette}}, \bibinfo {author} {\bibfnamefont {Sean~R.}\ \bibnamefont {O'Kelley}}, \bibinfo {author} {\bibfnamefont {Metin}\ \bibnamefont {Kayci}}, \bibinfo {author} {\bibfnamefont {Y.B.}\ \bibnamefont {Band}}, \bibinfo {author} {\bibfnamefont {Derek F.~Jackson}\ \bibnamefont {Kimball}}, \bibinfo {author} {\bibfnamefont {Alexander~O.}\ \bibnamefont {Sushkov}}, \ and\ \bibinfo {author} {\bibfnamefont {Dmitry}\ \bibnamefont {Budker}},\ }\bibfield  {title} {\enquote {\bibinfo {title} {Dynamics of a {{Ferromagnetic Particle Levitated}} over a {{Superconductor}}},}\ }\href {\doibase 10.1103/PhysRevApplied.11.044041} {\bibfield  {journal} {\bibinfo  {journal} {Physical Review Applied}\ }\textbf {\bibinfo {volume} {11}},\ \bibinfo {pages} {044041} (\bibinfo {year} {2019})}\BibitemShut {NoStop}%
\bibitem [{\citenamefont {Timberlake}\ \emph {et~al.}(2019)\citenamefont {Timberlake}, \citenamefont {Gasbarri}, \citenamefont {Vinante}, \citenamefont {Setter},\ and\ \citenamefont {Ulbricht}}]{timberlakeAccelerationSensingMagnetically2019}%
  \BibitemOpen
  \bibfield  {author} {\bibinfo {author} {\bibfnamefont {Chris}\ \bibnamefont {Timberlake}}, \bibinfo {author} {\bibfnamefont {Giulio}\ \bibnamefont {Gasbarri}}, \bibinfo {author} {\bibfnamefont {Andrea}\ \bibnamefont {Vinante}}, \bibinfo {author} {\bibfnamefont {Ashley}\ \bibnamefont {Setter}}, \ and\ \bibinfo {author} {\bibfnamefont {Hendrik}\ \bibnamefont {Ulbricht}},\ }\bibfield  {title} {\enquote {\bibinfo {title} {Acceleration sensing with magnetically levitated oscillators above a superconductor},}\ }\href {\doibase 10.1063/1.5129145} {\bibfield  {journal} {\bibinfo  {journal} {Applied Physics Letters}\ }\textbf {\bibinfo {volume} {115}},\ \bibinfo {pages} {224101} (\bibinfo {year} {2019})}\BibitemShut {NoStop}%
\bibitem [{\citenamefont {Ucar}(2021)}]{ucar2021polarity}%
  \BibitemOpen
  \bibfield  {author} {\bibinfo {author} {\bibfnamefont {Hamdi}\ \bibnamefont {Ucar}},\ }\bibfield  {title} {\enquote {\bibinfo {title} {Polarity free magnetic repulsion and magnetic bound state},}\ }\href@noop {} {\bibfield  {journal} {\bibinfo  {journal} {Symmetry}\ }\textbf {\bibinfo {volume} {13}},\ \bibinfo {pages} {442} (\bibinfo {year} {2021})}\BibitemShut {NoStop}%
\bibitem [{\citenamefont {Le~Lay}\ \emph {et~al.}(2024)\citenamefont {Le~Lay}, \citenamefont {Layani}, \citenamefont {Daerr}, \citenamefont {Berhanu}, \citenamefont {Dolbeault}, \citenamefont {Person}, \citenamefont {Roussille},\ and\ \citenamefont {Taberlet}}]{le2024magnetic}%
  \BibitemOpen
  \bibfield  {author} {\bibinfo {author} {\bibfnamefont {Gr{\'e}goire}\ \bibnamefont {Le~Lay}}, \bibinfo {author} {\bibfnamefont {Sarah}\ \bibnamefont {Layani}}, \bibinfo {author} {\bibfnamefont {Adrian}\ \bibnamefont {Daerr}}, \bibinfo {author} {\bibfnamefont {Michael}\ \bibnamefont {Berhanu}}, \bibinfo {author} {\bibfnamefont {R{\'e}my}\ \bibnamefont {Dolbeault}}, \bibinfo {author} {\bibfnamefont {Till}\ \bibnamefont {Person}}, \bibinfo {author} {\bibfnamefont {Hugo}\ \bibnamefont {Roussille}}, \ and\ \bibinfo {author} {\bibfnamefont {Nicolas}\ \bibnamefont {Taberlet}},\ }\bibfield  {title} {\enquote {\bibinfo {title} {Magnetic levitation in the field of a rotating dipole},}\ }\href@noop {} {\bibfield  {journal} {\bibinfo  {journal} {Physical Review E}\ }\textbf {\bibinfo {volume} {110}},\ \bibinfo {pages} {045003} (\bibinfo {year} {2024})}\BibitemShut {NoStop}%
\bibitem [{\citenamefont {Hermansen}\ \emph {et~al.}(2023)\citenamefont {Hermansen}, \citenamefont {Durhuus}, \citenamefont {Frandsen}, \citenamefont {Beleggia}, \citenamefont {Bahl},\ and\ \citenamefont {Bj{\o}rk}}]{hermansen2023magnetic}%
  \BibitemOpen
  \bibfield  {author} {\bibinfo {author} {\bibfnamefont {Joachim~Marco}\ \bibnamefont {Hermansen}}, \bibinfo {author} {\bibfnamefont {Frederik~Laust}\ \bibnamefont {Durhuus}}, \bibinfo {author} {\bibfnamefont {Cathrine}\ \bibnamefont {Frandsen}}, \bibinfo {author} {\bibfnamefont {Marco}\ \bibnamefont {Beleggia}}, \bibinfo {author} {\bibfnamefont {Christian~R.H.}\ \bibnamefont {Bahl}}, \ and\ \bibinfo {author} {\bibfnamefont {Rasmus}\ \bibnamefont {Bj{\o}rk}},\ }\bibfield  {title} {\enquote {\bibinfo {title} {Magnetic levitation by rotation},}\ }\href@noop {} {\bibfield  {journal} {\bibinfo  {journal} {Physical Review Applied}\ }\textbf {\bibinfo {volume} {20}},\ \bibinfo {pages} {044036} (\bibinfo {year} {2023})}\BibitemShut {NoStop}%
\bibitem [{\citenamefont {Hermansen}\ \emph {et~al.}(2025)\citenamefont {Hermansen}, \citenamefont {Durhuus},\ and\ \citenamefont {Bj{\o}rk}}]{hermansenMagneticLevitationLow2025}%
  \BibitemOpen
  \bibfield  {author} {\bibinfo {author} {\bibfnamefont {Joachim~Marco}\ \bibnamefont {Hermansen}}, \bibinfo {author} {\bibfnamefont {Frederik~Laust}\ \bibnamefont {Durhuus}}, \ and\ \bibinfo {author} {\bibfnamefont {Rasmus}\ \bibnamefont {Bj{\o}rk}},\ }\bibfield  {title} {\enquote {\bibinfo {title} {Magnetic levitation at low rotation frequencies using an on-axis magnetic field},}\ }\href {\doibase 10.1063/5.0267494} {\bibfield  {journal} {\bibinfo  {journal} {Applied Physics Letters}\ }\textbf {\bibinfo {volume} {127}},\ \bibinfo {pages} {012402} (\bibinfo {year} {2025})}\BibitemShut {NoStop}%
\bibitem [{\citenamefont {Seleznyova}\ \emph {et~al.}(2016)\citenamefont {Seleznyova}, \citenamefont {Strugatsky},\ and\ \citenamefont {Kliava}}]{seleznyova2016modelling}%
  \BibitemOpen
  \bibfield  {author} {\bibinfo {author} {\bibfnamefont {Kira}\ \bibnamefont {Seleznyova}}, \bibinfo {author} {\bibfnamefont {Mark}\ \bibnamefont {Strugatsky}}, \ and\ \bibinfo {author} {\bibfnamefont {Janis}\ \bibnamefont {Kliava}},\ }\bibfield  {title} {\enquote {\bibinfo {title} {Modelling the magnetic dipole},}\ }\href@noop {} {\bibfield  {journal} {\bibinfo  {journal} {European Journal of Physics}\ }\textbf {\bibinfo {volume} {37}},\ \bibinfo {pages} {025203} (\bibinfo {year} {2016})}\BibitemShut {NoStop}%
\bibitem [{\citenamefont {Greene}\ and\ \citenamefont {Karioris}(1971)}]{greene1971force}%
  \BibitemOpen
  \bibfield  {author} {\bibinfo {author} {\bibfnamefont {Jack~B.}\ \bibnamefont {Greene}}\ and\ \bibinfo {author} {\bibfnamefont {Frank~G.}\ \bibnamefont {Karioris}},\ }\bibfield  {title} {\enquote {\bibinfo {title} {Force on a magnetic dipole},}\ }\href@noop {} {\bibfield  {journal} {\bibinfo  {journal} {American Journal of Physics}\ }\textbf {\bibinfo {volume} {39}},\ \bibinfo {pages} {172--175} (\bibinfo {year} {1971})}\BibitemShut {NoStop}%
\bibitem [{\citenamefont {Ucar}(2022)}]{slowmo_ucar}%
  \BibitemOpen
  \bibfield  {author} {\bibinfo {author} {\bibfnamefont {Hamdi}\ \bibnamefont {Ucar}},\ }\href@noop {} {\enquote {\bibinfo {title} {Floating double ball magnets in slow motion (20220712 0652)},}\ }\bibinfo {howpublished} {\url{https://www.youtube.com/watch?v=cwW15lVrntQ}} (\bibinfo {year} {2022})\BibitemShut {NoStop}%
\bibitem [{\citenamefont {Schreckenberg}(2025)}]{slowmo_hugo}%
  \BibitemOpen
  \bibfield  {author} {\bibinfo {author} {\bibfnamefont {Hugo}\ \bibnamefont {Schreckenberg}},\ }\href@noop {} {\enquote {\bibinfo {title} {Magnetic levitation induced by a high speed rotating magnet, slow motion.}}\ }\bibinfo {howpublished} {\url{https://youtu.be/dqQcN_7UfrI}} (\bibinfo {year} {2025})\BibitemShut {NoStop}%
\bibitem [{\citenamefont {Richter}\ \emph {et~al.}(1996)\citenamefont {Richter}, \citenamefont {Dullin}, \citenamefont {Waalkens},\ and\ \citenamefont {Wiersig}}]{richter1996spherical}%
  \BibitemOpen
  \bibfield  {author} {\bibinfo {author} {\bibfnamefont {Peter~H}\ \bibnamefont {Richter}}, \bibinfo {author} {\bibfnamefont {Holger~R}\ \bibnamefont {Dullin}}, \bibinfo {author} {\bibfnamefont {Holger}\ \bibnamefont {Waalkens}}, \ and\ \bibinfo {author} {\bibfnamefont {Jan}\ \bibnamefont {Wiersig}},\ }\bibfield  {title} {\enquote {\bibinfo {title} {Spherical pendulum, actions, and spin},}\ }\href@noop {} {\bibfield  {journal} {\bibinfo  {journal} {The Journal of Physical Chemistry}\ }\textbf {\bibinfo {volume} {100}},\ \bibinfo {pages} {19124--19135} (\bibinfo {year} {1996})}\BibitemShut {NoStop}%
\end{thebibliography}%

\end{document}